\magnification\magstep1\hsize=16truecm \vsize=23truecm \baselineskip=0.7truecm
\def\({\left(} \def\){\right)}\def\[{\left[} \def\]{\right]}\def\ni{\noindent}
\font\bfb=cmbx12
\def\hb{\hfil\break}
\def\ni{\noindent}
\def\ve{\vfill\eject}
\overfullrule=0pt

\ni{\bfb Surface Incommensurate Structure in an Anisotropic Model\hb with 
competing 
interactions on Semiinfinite Triangular Lattice\dag}
\medskip\bigskip
\ni{\bf Pavol Pajersk\'y and Anton \v Surda}
\bigskip
\ni Institute of Physics, 
Slovak Academy of Science, D\'ubravsk\'a cesta 9, 842 28 Bratislava, Slovakia

\vfootnote{\dag}{E-mail address: fyzisurd@savba.sk}

\bigskip\bigskip

\ni{\bf Abstract.} An anisotropic spin model on a triangular semiinfinite 
lattice  with ferromagnetic nearest-neighbour interactions and
one antiferromagnetic next-nearest-neighbour interaction is investigated by 
the 
cluster transfer-matrix method. A phase 
diagram with $\langle2\rangle$ antiphase, ferromagnetic, incommensurate, and 
disordered 
phase is obtained. The bulk uniaxial incommensurate structure  modulated in 
the 
direction of the competing interactions is found between the  $\langle2\rangle$
antiphase and the disordered phase.
The incommensurate structure near the surface with
free and $\langle2\rangle$ boundary condition 
is studied at different temperatures. 
Paramagnetic damping at the surface and  enhancement  of 
the incommensurate structure in the 
subsurface region at high temperatures and a new subsurface 
incommensurate structure 
modulated in two directions at low temperatures are found.

\ve


\noindent{\bf1. Introduction}
\bigskip

The cluster transfer-matrix method was found as a useful tool for description 
of  commensurate  and incommensurate  structures in two dimensional and 
three-dimensional spin lattice models. It is able to describe  floating 
incommensurate structures in two dimensions[1, 2, 3] as well as an infinite 
number 
of commensurate structures in three-dimensional models [4]. The method yields 
phase 
diagram of the model, free energy,   correlation functions, and magnetization
as a function of 
coordinates. As all the calculations are performed in real space, there is a 
possibility to study the properties of spatially inhomogeneous systems, e.g. a 
lattice  with a surface, where the inhomogeneity is localized in one direction,
and it is homogeneous in the others.

The cluster transfer-matrix method is a generalized mean field approximation 
-- it uses auxiliary effective multisite fields that are not directly related 
to the magnetization or multisite correlation functions of the model, 
nevertheless, the correlation functions  can be calculated from them. 
In two dimensions, the spatial dependence of the fields in one direction is 
obtained 
by simple iterating the effective fields from one lattice row  perpendicular 
to it, to the following one. It is more difficult to get the spatial 
dependence inside the row, i.e. in the direction perpendicular to the 
direction of the iteration. Here, the  correlation functions of a row
of spins interacting by original interactions of the Hamiltonian plus by the 
spatially dependent  effective multisite fields
should be found. For that reason 
the iteration in the systems with uniaxial incommensurate structure is always 
performed in the direction of the incommensurate modulation.  

The transfer matrix formalism is also used in  derivation of the 
fermion Hamiltonian in the domain wall theory of commensurate-incommensurate 
(C-IC) transitions in 2D lattice models [5]. The 
domain walls in the incommensurate  structure are described by world lines of 
fermions and therefore, 
the transfer   matrices are defined on columns of sites   in the direction of 
the incommensurate  modulation, i.e.  perpendicular to the transfer matrices 
used in our method.

It is simple to study 
surface or subsurface properties in the systems where the  surface is 
perpendicular to the direction of the iteration. In fact, this is done 
always when the bulk properties of the system are calculated, as the starting 
values of the effective fields in the iteration procedure play a role of 
 surface boundary conditions. In this case,  the most conspicuous 
properties of the subsurface region appear  for paramagnetic phase
near the phase transition line with the incommensurate or ferromagnetic 
structure. In the first case, the surface effects attenuate with 
distance 
from the surface in an oscillatory way, in the latter case, monotonically.

In this paper we study surface and subsurface properties of a two dimensional 
 system 
with the surface orientated {\it parallel} to the incommensurate modulation. 
Now, the 
effective fields used in the iteration procedure are  functions of distance 
from the surface and all of them  should be stored in the computer memory.  
Fortunately, 
far enough from the critical point the surface effects are confined to a 
relatively narrow region, outside  which the effective fields acquire constant
bulk values. Thus, the shortest possible distance from the critical point in 
the parameter space is limited  by  computer 
memory in our calculations.

The cluster transfer-matrix method is related to the mean field approximation 
of Jensen and Bak [6]  were a nonlinear mapping of site magnetizations instead 
of effective fields is carried out. There,  in distinction to our method, the 
physically stable solutions are mathematically unstable. 
The exact nonlinear mapping is possible on lattices without  closed
loops, like Cayley tree and Bethe lattice. These lattices  
are characterized  by the site coordination number rather 
than the dimensionality. This 
nonlinear mapping technique were applied to various models including Potts 
[7] Ising [8] and ANNNI model [9]. It is difficult to relate the results for 
the hierarchical  lattices 
to those for Bravais lattices. Nevertheless, the phase diagram of the ANNNI 
model on Cayley tree with infinite coordination number [9] bear similar 
features  to that of the 3D ANNNI model.

The ANNNI 
model defined on the square lattice and anisotropic antiferromagnetic (AA) 
model on the  triangular 
lattice are the most simple 2D models displaying an uniaxial incommensurate 
structure. 
There are two competing interactions in the both models: ferromagnetic nn and 
and 
antiferromagnetic third-nearest neighbour interactions in the ANNNI model and 
antiferromagnetic nn and ferromagnetic nnn in the AA model on the triangular 
lattice. Both models are anisotropic, i.e. two of the  third-nearest neighbour 
interactions and one or two of the nnn interactions are missing. Both models 
were investigated by the cluster transfer-matrix approximation and the results 
were consistent with numerous other approaches  like Monte Carlo calculations, 
series expansions, cluster variation method,  domain wall theory, finite-size 
scaling [10, 11, 12, 13].

The 
phase diagrams of 2D models are more simple than those in the 3D case. They 
consist  of a small  number of  commensurate phases and a single region of 
a floating incommensurate phase. In the ANNNI model, the nn interactions are 
ferromagnetic 
and consequently, the rows are ordered ferromagnetically; in the AA model they 
have a commensurate structure with periodicity of three lattice constants.

Here, we study a natural generalization of the ANNNI model to the triangular 
lattice. The nn interactions are ferromagnetic and one nnn interaction 
instead of third-next nearest interaction is 
antiferromagnetic, i.e, the signs of the interactions are opposite to the 
above 
described AA model. The phase diagram and all  other properties are similar to 
those in the ANNNI model. Hence, it can be expected that the surface effects 
in the ANNNI model on the square lattice are closely related to the ones 
described bellow.

\bigskip\medskip
\noindent{\bf 2. Model and  method}
\bigskip
We consider an anisotropic ferromagnetic model of Ising spins ($\sigma=\pm 1$)
on a triangular semiinfinite lattice
interacting by nearest neighbour (nn) and one next-nearest-neighbour  (nnn) 
interactions. All the nn interactions of the model are ferromagnetic.
Two of the three  possible nnn interactions are missing and the remaining one 
is antiferromagnetic.  
The  triangular 
lattice with the spin interactions and the clusters used in further 
calculation are shown 
in Fig.~1, where $j=0,\dots, \infty$ and $i=-\infty,\dots, \infty$.
We shall calculate the free energy and  the local 
magnetization by the cluster-matrix method developed by one of 
us [14]. 

The cluster-matrix method is based on a subsequent summation of the weight
functions $\exp[\beta H(\sigma _i)]$ over the spin variables
in the consecutive rows when
calculating the partition function. For the computational reasons
zigzag rows shown in Fig.~1 perpendicular to the nnn interaction are chosen. 
The lattice surface is perpendicular to the rows and the expected direction of 
the domain walls. It is put at the column $j=0$.

Let us write the Hamiltonian of the model 
$$
H=\sum J_1 \sigma _{i,j}(\sigma _{i,j+1} +\sigma _{i,j-1}+\sigma _{i,j+2})
+J_2\sigma _{i,j}\sigma _{i+1,j}
     $$
as a sum of energies of the $2\times 4$ clusters
$$
H=\sum_{i,j}[G_{i,2j}+G'_{i,2j+1}]
$$

We use two types of the $2\times 4$ clusters, that are shifted by one
lattice constant and can be transformed into each other by translation
and rotation by an angle $180^{\circ}$ in the plane of the lattice. 
They are shown in the inset of Fig.~1.

The energy of the bulk cluster of the first type is 
$$
\eqalign{
G_{i,2j}  = J_1& \Bigl[\sigma _{i,2j+1}\({\sigma _{i,2j}\over6} + 
{\sigma _{i,2j+2}\over6} +{\sigma _{i,2j+3}\over3} +{\sigma _{i+1,2j}\over3}+
{\sigma _{i+1,2j+2}\over3}
\) \cr
+& \sigma _{i+1,2j+2}\({\sigma _{i+1,2j+1}\over6} +{\sigma _{i+1,2j+3}\over6} 
+ {\sigma _{i+1,2j}\over3} +{\sigma _{i,2j+3}\over3}\)\cr
+& {\sigma _{i,2j+2}\sigma _{i,2j+3}\over6}+ 
{\sigma_{i+1,2j}\sigma _{i+1,2j+1}\over6}\Bigr]+
J_2\Bigl[{\sigma _{i,2j}\sigma _{i+1,2j}\over4} \cr
+ &{\sigma _{i,2j+1}\sigma _{i+1,2j+1}\over4}+
{\sigma _{i,2j+2}\sigma _{i+1,2j+2}\over4} +
{\sigma _{i,2j+3}\sigma _{i+1,2j+3}\over4}\Bigr]  ,\cr}                 
\eqno(1)
$$
where $J_1$ is the nearest-neighbor interaction and $J_2$ is the 
next-nearest one and $j=1,2,\dots, \infty$. The terms in (1) are divided by 
the number of appearances of the particular bond in different overlapping 
clusters. The expression for the energy of the cluster of the second type 
denoted by $G'_{i,2j+1}$
can be found by interchanging $i\leftrightarrow i+1$ and $2j\rightarrow 2j+1$
at the right hand side of (1). The denominators in (1) are different in the 
expressions for the energies  $G_{i,0}$ and $G'_{i,1}$ of the surface 
clusters, because the translational symmetry is broken here.

The evaluation of the partition function
$$
Z=\sum_{\{\sigma _i\}} \exp[\beta H(\sigma _i)]
$$
can be transformed to the calculation of the numbers $\lambda _i$ appearing
as normalization factors in the iterative procedure for auxiliary functions 
$\Psi_i$ 
$$
\sum_{S_i}\Psi_i(S_i)T_i(S_i,S_{i+1}) = \lambda _i 
\Psi_{i+1}(S_{i+1})                                     \eqno(2)
$$
starting from an appropriate function $\Psi_1(S_1)$[1, 2, 3].
($S_i$ denotes a row 
variable $S_i\equiv\{\sigma _{i,0}\dots,\sigma _{i,2j},\sigma _{i,2j+1},\sigma 
_{i,2j+2}\dots
\}$  
and $T_i(S_i,S_{i+1})= \exp\[\beta \sum_{j=0}^{j=+\infty} 
(G_{i,2j}+G'_{i,2j+1})\]$.) $Z=\prod_{i=0}^\infty \lambda_i$.

Unfortunately, each of the auxiliary functions $\Psi_i(S_i)$ acquires
an infinite number of values and an approximation should be done to perform the
summation on the left hand side of (2).
                                                                               
Assuming an asymptotic behaviour of correlation functions already at distances 
exceeding 
 the cluster size, we can try to factorize $\Psi_i(S_i)$ in the same way as 
the function $T_i(S_i,S_{i+1})=\prod_{j=0}^\infty \exp(G_{i,2j})
\exp(G'_{i,2j+1})$
i.e. 
$$
\Psi_i(S_i)\simeq\prod_{j=0}^\infty
\Psi_{i,2j}(s_{i,2j}^k)\Psi'_{i,2j+1}(s_{i,2j+1}^k)
                                                               \eqno(3)
$$
where $s_{i,l}^k$ denotes a set of site variables of a finite row cluster
$s_{i,l}^k=(\sigma _{i,l},\dots, \sigma _{i,l+k})$ and 
$\Psi_{i,2j}(s_{i,2j}^k),\ \Psi'_{i,2j+1}(s_{i,2j+1}^k)$ are the cluster   
auxiliary functions acquiring a finite number of values.

The number $k$ characterizes the order of the approximation and was taken 
equal 
to 4 what is the width of the clusters in (1). (Fig.~1) 
 
The logarithms of the values of the cluster auxiliary function for different 
cluster configuration represent the above mentioned multisite effective 
fields. As the functions are defined  on finite clusters, only short-range 
effective interactions are taken into account in our approximation.

Substituting (3) into (2), we obtain a relation between the known functions 
$\Psi_{i,2j}$, $\Psi'_{i,2j+1}$  found in the preceding iteration step and 
the new functions $\Psi_{i+1,2j}$, $\Psi'_{i+1,2j+1}$.
The expression for $\Psi_{i+1,2j}$, $\Psi'_{i+1,2j+1}$ in terms of
$\Psi_{i,2j}$, $\Psi'_{i,2j+1}$ can be found by a partial summation of
the both sides of (2). This problem is one-dimensional and the partial 
summation can be  done exactly---again by the technique of  auxiliary 
functions   as shown in detail in previous papers [1, 2, 3, 14]. 
In contrast with the previous calculations on infinite lattices,  the
equation (2) has no translational symmetry in the row direction due to
the presence of the surface and the equation should be solved for all
$\Psi_{i+1,2j}$, $\Psi'_{i+1,2j+1}$, $j=0,\dots,\infty.$ In practice, the 
cluster auxiliary functions converge to their bulk values fast if we are far 
enough from the continuous incommensurate-commensurate (IC-C) phase 
transition. 
We confined ourselves to the distances from the IC-C phase transition line 
where the number of the cluster auxiliary functions taken into account did not 
exceed 
$j=400.$

In the paramagnetic and ferromagnetic phase $\Psi_{i,2j}$, $\Psi'_{i,2j+1},$ 
do not depend on $i$, 
in the $\langle2\rangle$ antiphase consisting of zig-zag rows with alternating 
magnetization,
$\Psi_{i,2j}$, $\Psi'_{i,2j+1}$ are periodic functions of 
$i$ with  
period of two. In the incommensurate  phase, their period 
is a continuous function of the interaction constants.
The functions $\Psi_{i,2j}$, $\Psi'_{i,2j+1}$ are $j$ independent in the bulk,
i.e. the row structure is ferromagnetic or paramagnetic in all phases.
Nevertheless, they are strongly   spatially  modulated near the surface what 
leads even to the  areas  of reversed magnetization.

From the knowledge of the auxiliary functions $\Psi$ and $\Psi' $, it is 
possible to find the site magnetizations. We have 
$$
\langle\sigma _{i,l}\rangle= \sum_{S_{i}} \prod_j 
\Psi_{i,2j}(s_{i,2j}^k) \Psi'_{i,2j+1}(s_{i,2j+1}^k) \sigma _{i,l}
\tilde\Psi_{i,2j}(s_{i,2j}^k) \tilde\Psi'_{i,2j+1}(s_{i,2j+1}^k).
                                                     \eqno(4)
$$
where $\tilde\Psi',\ \tilde\Psi'$ are the functions that  are
calculated by the same iteration  procedure as in (2) but in the 
opposite direction.


\bigskip\medskip
\ni{\bf3. Results and discussion}
\bigskip

The calculations have shown that the anisotropy model with competing nn 
and nnn interactions on a triangular
lattice can be found in one of the four phases: disordered  paramagnetic,
commensurate $\langle2\rangle$ antiphase, ferromagnetic, and the 
incommensurate  one lying between them. 

The phase diagram of the model,  shown in Fig.~2,
is similar to the phase diagram of the 2D ANNNI model on the
square lattice [2, 10, 11].
Near the multiphase point $J_1/J_2=1$ the disordered phase should persist 
to $T=0$ in the form of an extremely narrow
strip. Unfortunately, by our method it is not possible to
verify this fact at very low temperatures and the direct
phase transition between the ferromagnetic and the incommensurate
phase cannot be excluded. On the other hand, we found no signs confirming the 
opposite case.
For $J_1/J_2\rightarrow 0$ the incommensurate 
phase seems to be  stable down to the point $T=0$.
The IC-disorder phase transition line seemingly tends to a Lifshitz point
at the ferro-disorder phase transition line but at the close vicinity of it, 
it turns abruptly down and apparently meets it  at $T=0$. It is seen that a 
less 
careful numerical treatment of the problem could lead to an erroneous 
confirmation of the Lifshitz point in 2D ANNNI model.

When we put $J_2=0$, the exactly solvable ferromagnetic Ising model on the 
triangular lattice  with  the critical temperature  $T_c=3.732\dots$ is 
restored. 
Our method yields $T_c=3.64$. We believe that the comparison of this two 
values suggests  the accuracy of the whole phase diagram shown in Fig.~2.
 
The interaction constants and temperature in 
all further presented results are localized 
 in the areas  denoted by two short bars in the incommensurate region of the 
phase diagram
near the phase transition lines with the disordered paramagnetic phase and 
$\langle2\rangle$ antiphase, respectively.
At the higher temperature the bulk magnetization is of a sinusoidal shape. 
At the lower temperatures the structure consists of strip-like 
$\langle2\rangle$ antiphase domains. Their width is growing to infinity with 
temperature approaching the IC-C phase transition line. The bulk structures 
can be seen   in the depth more then 400 columns from 
the surface in the following figures.

We consider two different boundary
conditions at the surface: the free boundary condition (FBC) and 
the $\langle2\rangle$ antiphase boundary condition (ABC).
In our approach the boundary condition is given by the starting values 
$\Psi_{i,0}$ of the auxiliary function.
For FBC all values of the  auxiliary function  on the surface
are taken equal to unity. 
The ABC  boundary conditions can be  simulated by the values of the cluster 
auxiliary function deep  in the bulk of  the $\langle2\rangle$ structure at low
temperature.
Actually, they have been taken as an output of calculation at
$J_1/J_2=0.5, T/J_2=0.1$ for
$j=600$.

The site magnetizations  at every second zig-zag row and at first 480 
subsurface columns 
for $T/J_2=1.47, 1.252, 1.247, 1.241$ are shown in Fig.~3a-d. 
All these figures are calculated for the FBC. 
In Fig.~4., the magnetization along the rows of sites with maximum absolute 
value of magnetization as a function of distance from the surface is shown.

As expected, the amplitude of the sinusoidal  magnetization  at the 
surface is diminished by the absence of interactions for FBC at the 
temperature $T/J_2=1.47$, close to the paramagnetic structure. This  
suppression is replaced by an enhancement of the incommensurate waves of 
magnetization in the narrow subsurface region (Fig.~3a).  The presence of the 
surface affects, approximately, only first 60 columns at this temperature.
Similar increase of the magnetization profile near the surface was found for 
semi-infinite ferromagnetic Ising model [15].

The situation is different for temperatures  near the IC-C phase transition 
where wide one-dimensional domains 
of $\langle2\rangle$ 
structure bounded by domain walls perpendicular to the nnn  $J_2$ interaction 
$J_2$ occur in the bulk. Two neighbouring 
domains differ by a phase shift of $\pi$ (Fig.~3b--3d). Near the surface, 
the strip-like bulk domains  become modulated, as well, and incommensurate 
domains 
are formed in the direction perpendicular of the nnn interaction. By 
approaching 
the IC-C phase transition line (lowering the temperature) the region of the 
biaxial incommensurate structure becomes wider and its depth 
changes linearly with temperature, as shown in Fig.~5. Extrapolating the 
linear plot to the temperature of the phase transition 
$T_{\rm IC-C}=1.1876$, the width of the biaxially modulated structure at 
the critical point is found approximately equal to  1300 columns.

Fig.~6 and 7 shows that the influence of the boundary condition is small. The 
change from FBC  to ABC affects only first few subsurface columns. The phase of
the $\langle2\rangle$  structure is fixed at the surface, but it does 
not influence the phase in the bulk that changes quite freely at the domain 
walls.
The structure perpendicular to the surface is unaffected, as well.

The cluster auxiliary 
functions $\Psi_{i,2j}$, $\Psi'_{i,2j+1}$ 
are
the direct output of the iteration procedure 
and in the bulk behave similarly 
to the magnetization shown in the previous figures.  In the low-temperature 
incommensurate structure, they form one-dimensional strip-like domains 
possessing the 
symmetry of $\langle2\rangle$ phase. Near the surface the domains bend in the 
 direction opposite to the direction of the iteration.

This situation is shown in Fig.~8, where one of the 64 values
of the cluster auxiliary function $\Psi_{i,2j}$ at subsurface lattice sites is 
plotted. The direction of the  iteration is from the left to the right, i.e. 
the domains a bended backwards. It looks like there is a friction between the 
auxiliary-function  structures and the surface when the space evolution of the 
auxiliary function is calculated by the iteration procedure.

The domains are bended    but near the surface they are again straight. The 
bending 
angle between the direction of the bulk and surface domain is increasing  when 
approaching the critical line as shown in Fig~5. At the critical temperature, 
the angle 
tends to 45$^\circ$. 

The magnetization 
is calculated from  eq.~4 which involves two auxiliary function  which are 
iterated in opposite directions and therefore their surface parts are bended 
in the opposite sense. 
The 
surface incommensurate structure  
is in fact their  interference pattern and the resulting 
modulation of the magnetization has a two-dimensional character.

As the the bending angle of the auxiliary-function domains is between 0$^\circ$
and 45$^\circ$
the  2D magnetization domains are oblong at  higher temperatures and become 
square-like  near the phase transition line. On the other hand, if the linear 
extrapolation is applicable up to the critical line, the width of the bulk 
domains becomes infinite while  the width of the surface region remains 
finite. Thus,  the subsurface structure should in fact disappear at the phase 
transition. 

The width of subsurface structure was measured as a distance from surface to 
the point of the maximum curvature of the bended domain wall.

In conclusion,  influence of the surface on the incommensurate structure of 
an anisotropic Ising model on a triangular lattice was investigated by 
the cluster 
transfer-matrix method. The uniaxial incommensurate structure in a finite 
region 
near the surface changes its character and becomes biaxial. The width of the
biaxially modulated structure seems to be finite at critical temperature. 
Formation of the biaxial structure can be interpreted as an 
interference pattern of two auxiliary-function wave-like structures. 

From a more physical point of view, 
it can be seen as a result of a misfit between the bulk structure
with longer periodicity than that of the surface structure due to the absence 
of a part of  interactions at the surface that is equivalent to an effective 
increase of temperature.

\vfil\eject

\ni{\bf REFERENCES}
\bigskip
\item{[1]} \v Surda A 1991 {\sl Physica A} {\bf 178} 332
\item{[2]}Karasov\'a I. and \v Surda A.~1993 {\sl J. Stat. Phys.} {\bf 70} 675 
\item{[3]}Pajersk\'y P. and  \v Surda A.~1994 {\sl J. Stat. Phys.} {\bf 76} 1467
\item{[4]} \v Surda A 1996 {\sl Commensurate and incommensurate structures in 3D 
ANNNI model} (to appear)
\item{[5]}den Nijs M 1992 {\sl Phase Transition and Critical Phenomena} Vol~12
(New York: Academic Press)
\item{[6]}Jensen M H, Bak P 1983 {\sl Phys. Rev. B} {\bf 27} 6853
\item{[7]} Monroe J L 1996 
{\it J.~Phys. A: Math. Gen.} {\bf 29} 5421
\item{[8]} M\'elin R, Angl\`es d'Auriac J C, Chandra P and  Dou\c cot B 1996
{\it J.~Phys. A: Math. Gen.} {\bf 29} 5773
\item{[9]}Yokoi C S~O, de Oliveira M J and Salinas S~R 1985 {\sl Phys. Rev. 
Lett.} {\bf 54} 163
\item{[10]}Selke W 1988 {\sl Phys. Rep.} {\bf 170}  587
\item{[11]}Selke W 1992 {\sl Phase Transition and Critical Phenomena} Vol~15 
(New York: Academic Press)
\item{[12]} de Queiroz S~L A and Domany E 1995 {\sl Search for 
Kosterlitz-Thouless 
transition in a triangular Ising antiferromagnet with further-neighbour 
ferromagnetic interactions} SISSA-preprint cond-mat 9507076
\item{[13]}Kitatani H and Oguchi T.~1988 {\sl J.~Phys. Soc. Jpn.} {\bf 1344}  
\item{[14]}\v Surda A 1991 {\sl Phys. Rev. B} {\bf 43} 908 
\item{[15]}Czerner P and Ritschel U  1996 {\sl Magnetization  profile in the 
$d=2$ 
semi-infinite Ising model and crossover between ordinary and normal transition}
SISSA-preprint cond-mat 969120

\vfil\eject
\ni{\bf FIGURE CAPTIONS}
\bigskip

\ni Fig. 1. Anisotropic model on  triangular half lattice with two competing 
interactions. Each spin interacts with 6 nn spins by $J_1$ interaction 
(thin lines) and 2 nnn spins by   $J_2$  interaction (thick lines).
The two types $2 \times 4$ clusters used in the calculation are in the inset.
Our zig-zag row encompasses  two ordinary rows of the triangular lattice.
\bigskip

\ni Fig. 2. Phase diagram of the model.
Two short bars (at $J_1/J_2=0.5$, $T/J_2=1.241, 1.47$) show the parameter 
regions of the  calculations presented in Figs.~3--8.

 \bigskip

\ni Fig. 3a. 
Site magnetizations $m_{i,j}=\langle\sigma_{i,j} \rangle$ 
at the  first 480 subsurface columns $(j=1,\dots, 480)$  for 
$T/J_2=1.47, J_1/J_2=0.5$ and FBC. 
\bigskip

\ni Fig. 3b. 
Site magnetizations $m_{i,j}$ 
at the  first 480 subsurface columns  for 
 $T/J_2=1.252, J_1/J_2=0.5$ and FBC. 
\bigskip

\ni Fig. 3c. 
Site magnetizations $m_{i,j}$ 
at the  first 480 subsurface columns  for 
$T/J_2=1.247, J_1/J_2=0.5$ and FBC. 
\bigskip

\ni Fig. 3d. 
Site magnetizations $m_{i,j}$ 
at the  first 480 subsurface columns  for 
$T/J_2=1.241, J_1/J_2=0.5$ and FBC. 
\bigskip

\ni Fig. 4. 
Site magnetizations $m_{i,j}$ in the  direction perpendicular to the 
surface as a function of 
the column number  $j$  at $T/J_2=1.241$ (the thickest curve), 1.247, 1.252, 
1.47 (the 
thinnest curve) and $J_1/J_2=0.5$ for 
{FBC}. 
The curves follow one of ridges of the structures in  Figs.~3a--d.
\bigskip

\ni Fig.~5. 
Width of the surface affected region (stars) and angle $\alpha $ between  
of the bulk  auxiliary-function domain wall and the domain wall near the 
surface (triangles). 

\bigskip

\ni Fig. 6. 
Site magnetizations $m_{i,j}$ 
at the first 480 subsurface columns  for 
$T/J_2=1.241, J_1/J_2=0.5$ and ABC. 
\bigskip

\ni Fig. 7. 
Site magnetizations $m_{i,j}$ in the  direction perpendicular to the 
surface as a function of 
the column number  $j$ at $T/J_2=1.241$ (the thickest curve), 1.247, 1.252, 
1.47 (the 
thinnest curve) and $J_1/J_2=0.5$ for 
{ABC}. 
The curves follow one of the ridges of the structure in  Fig.~6.
\bigskip

\ni Fig. 8.  Cluster auxiliary function $\Psi_{i,j}$  
at the first 480 subsurface columns  for
 $T/J_2=1.247, J_1/J_2=0.5$ and FBC. Direction of iteration is from the 
left to the right.
\bigskip

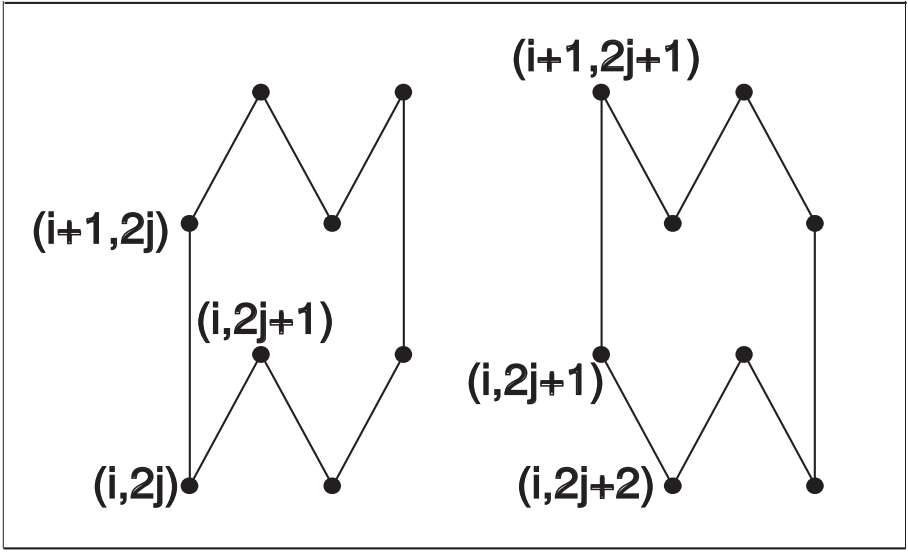
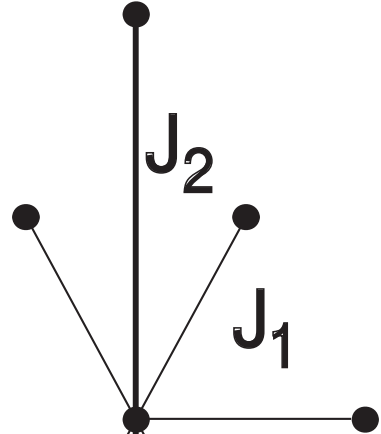
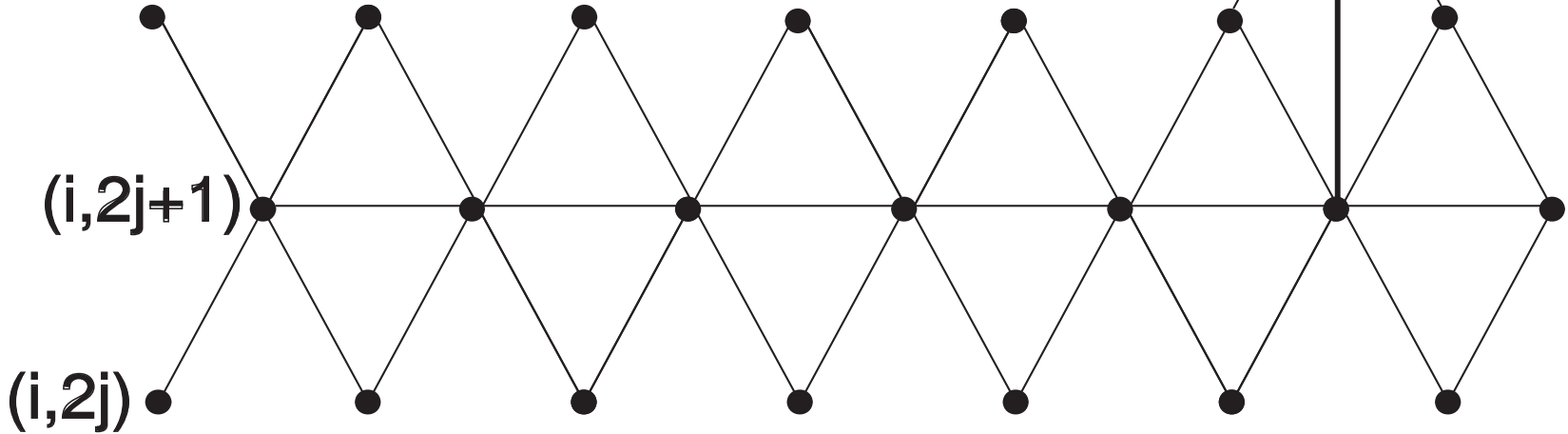
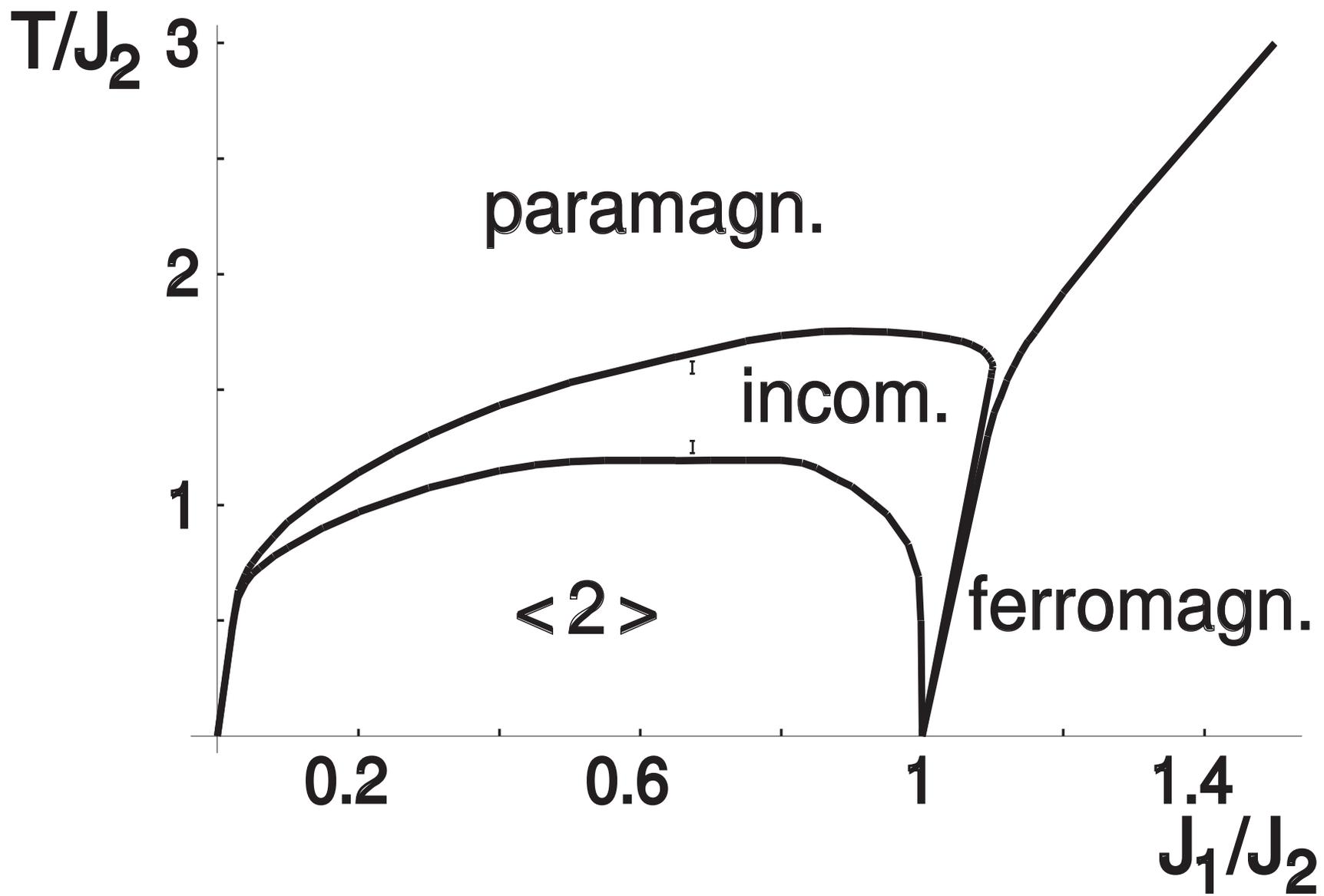
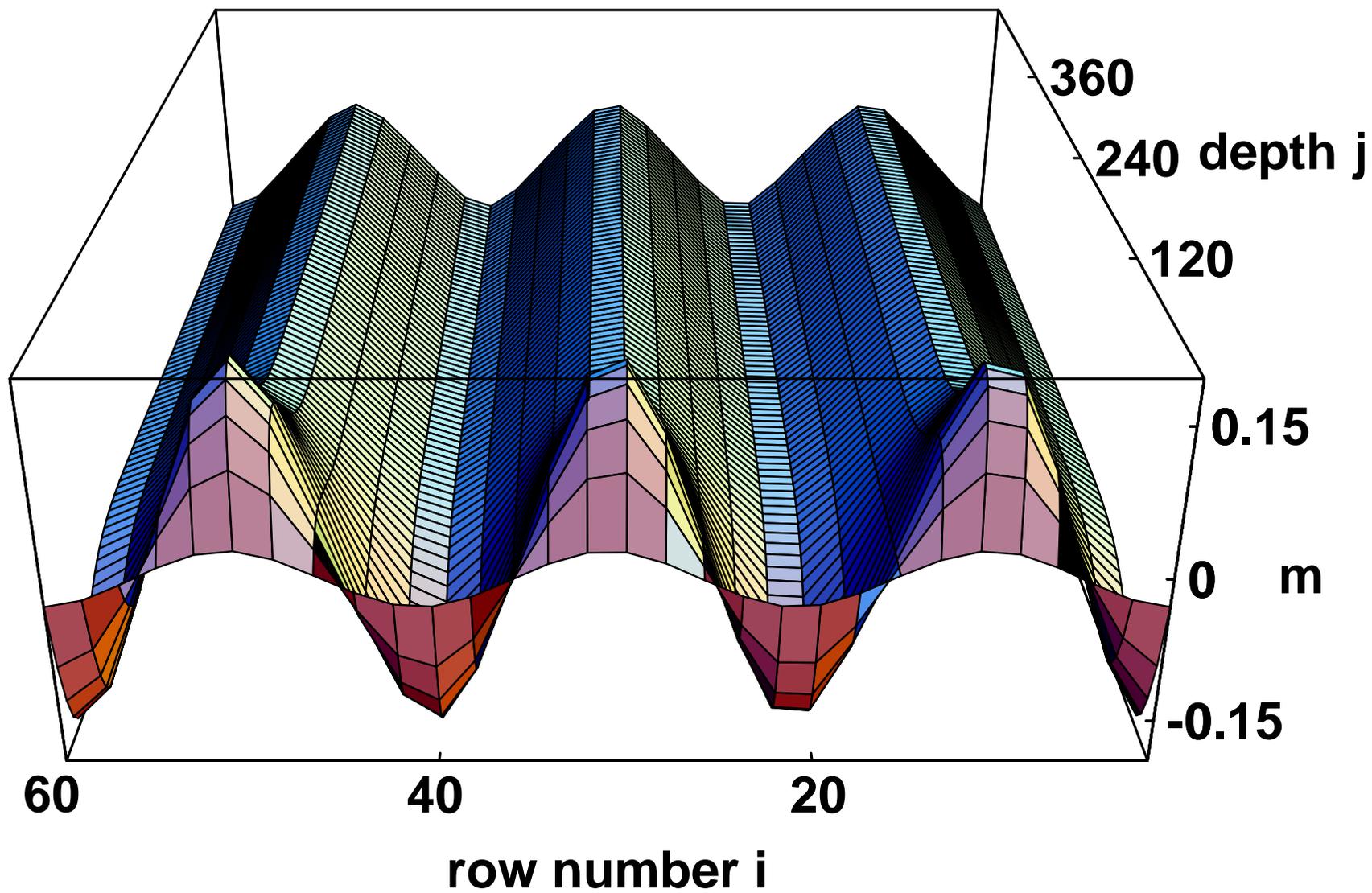
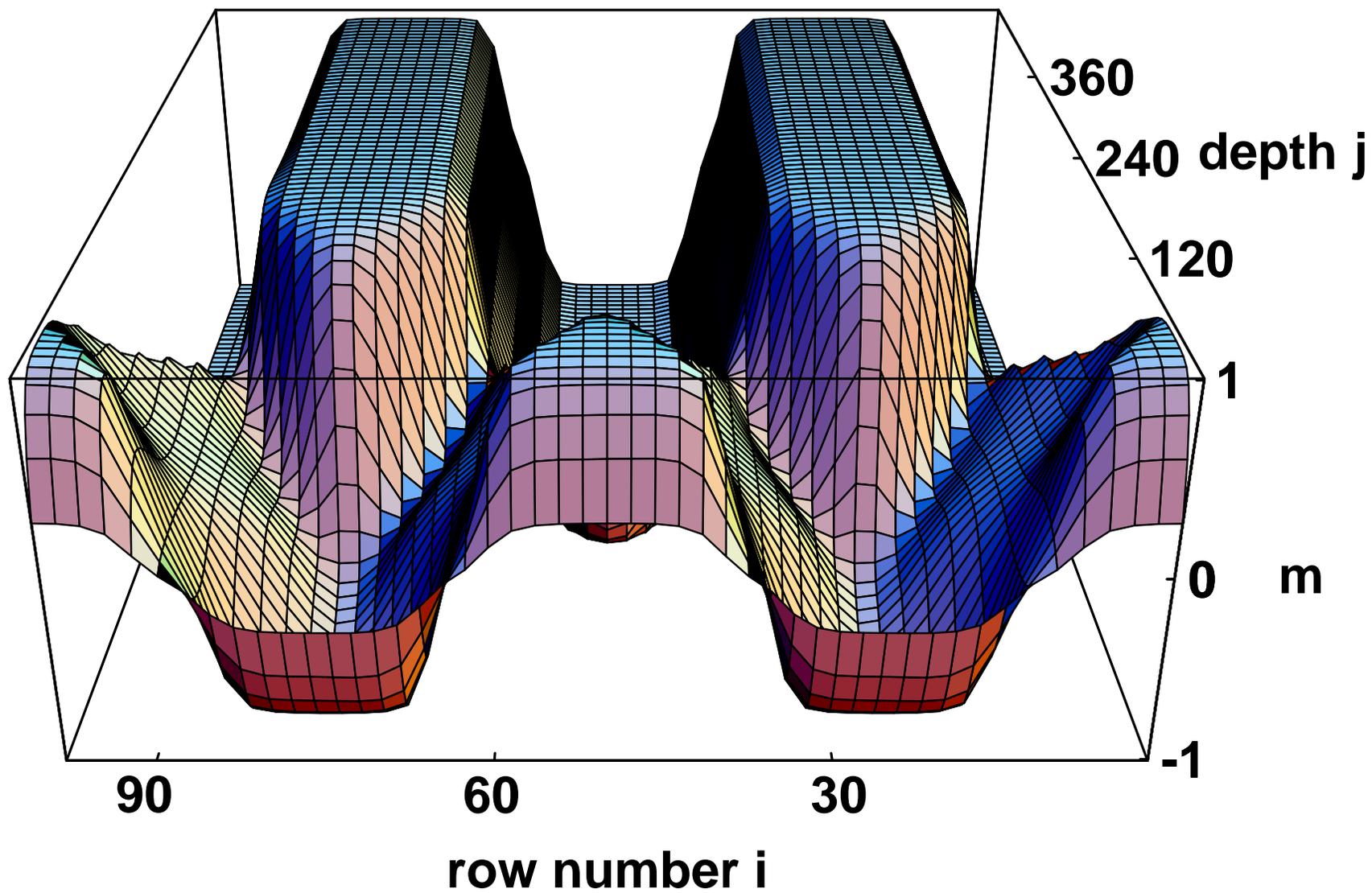
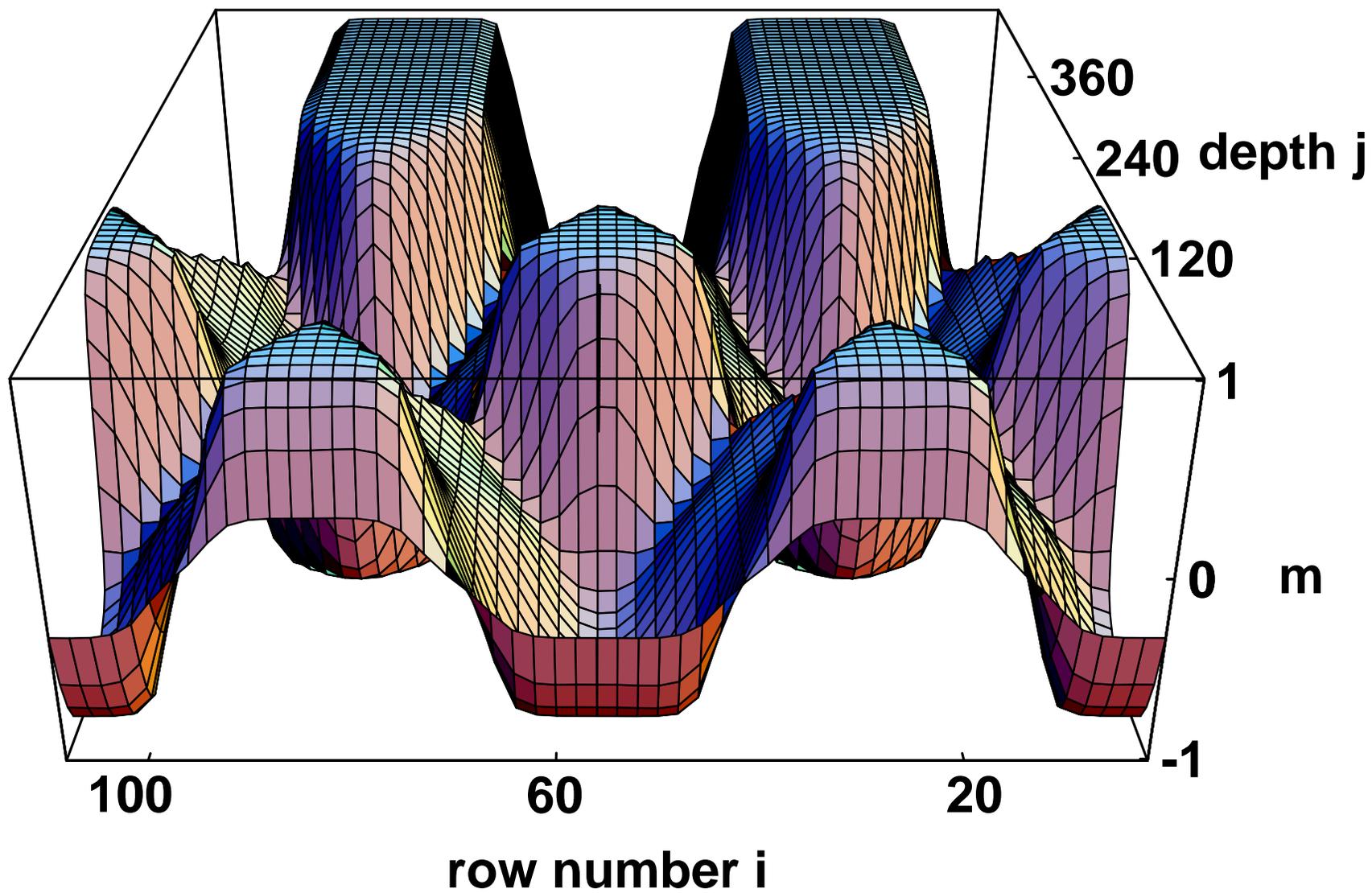
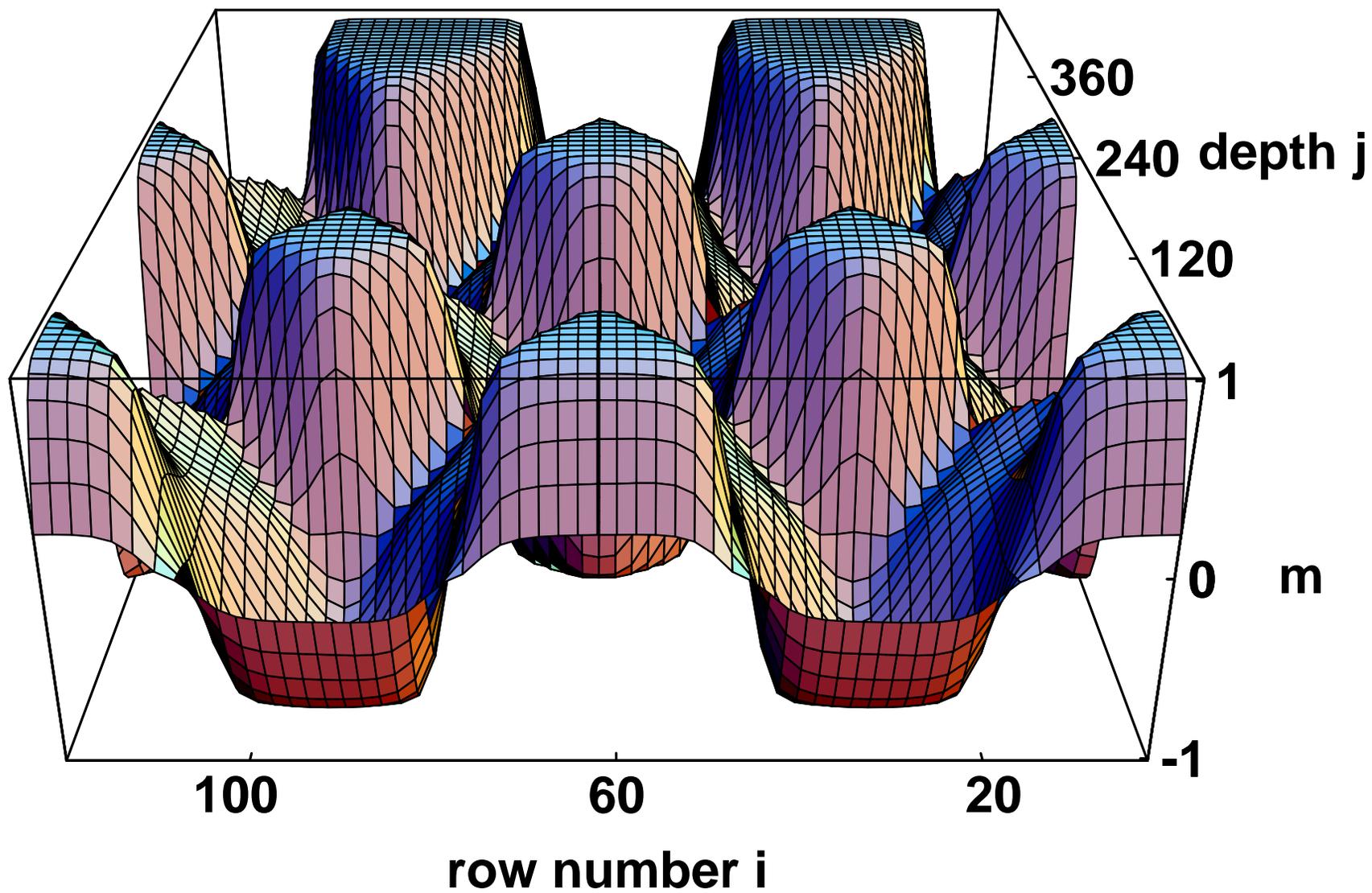
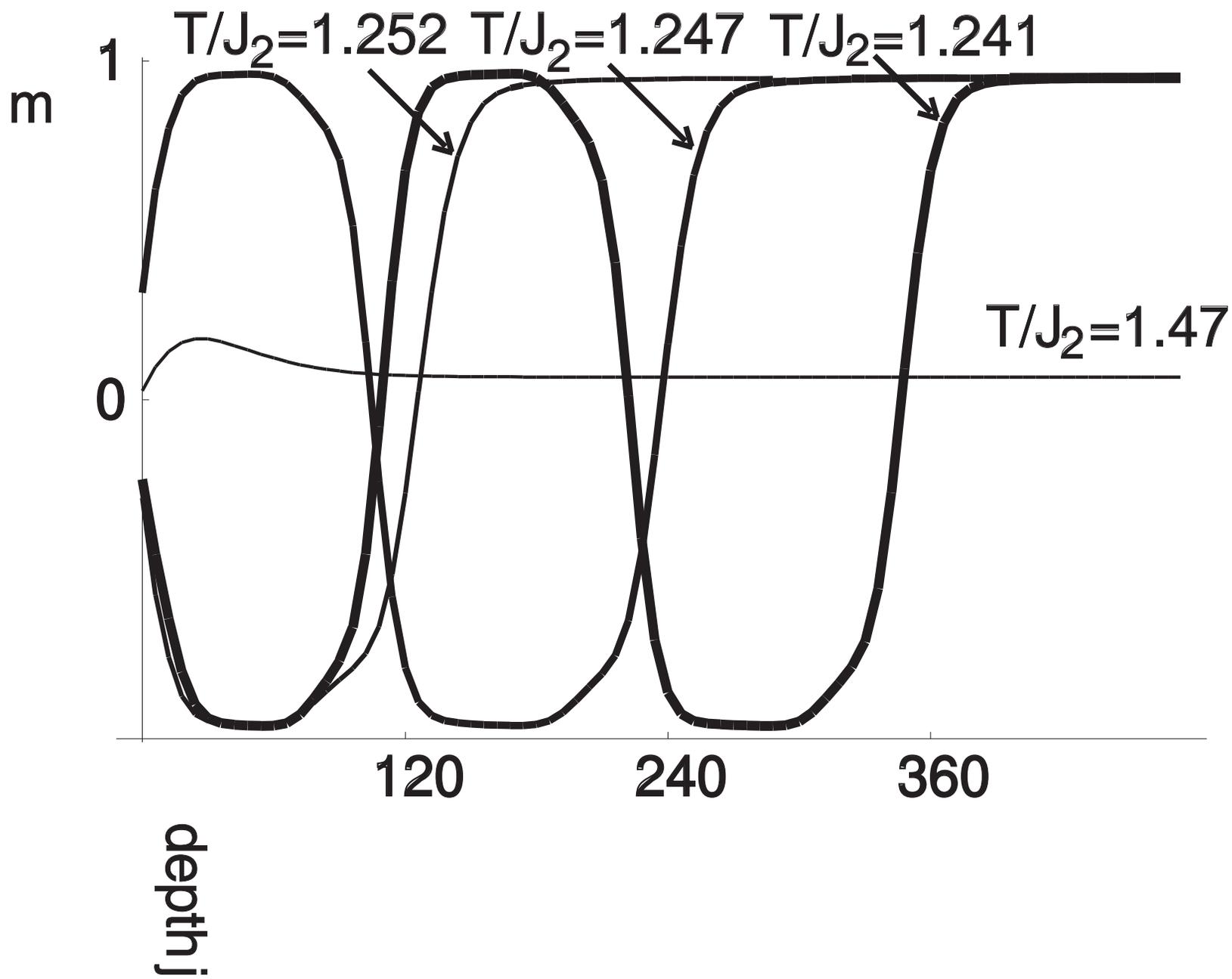
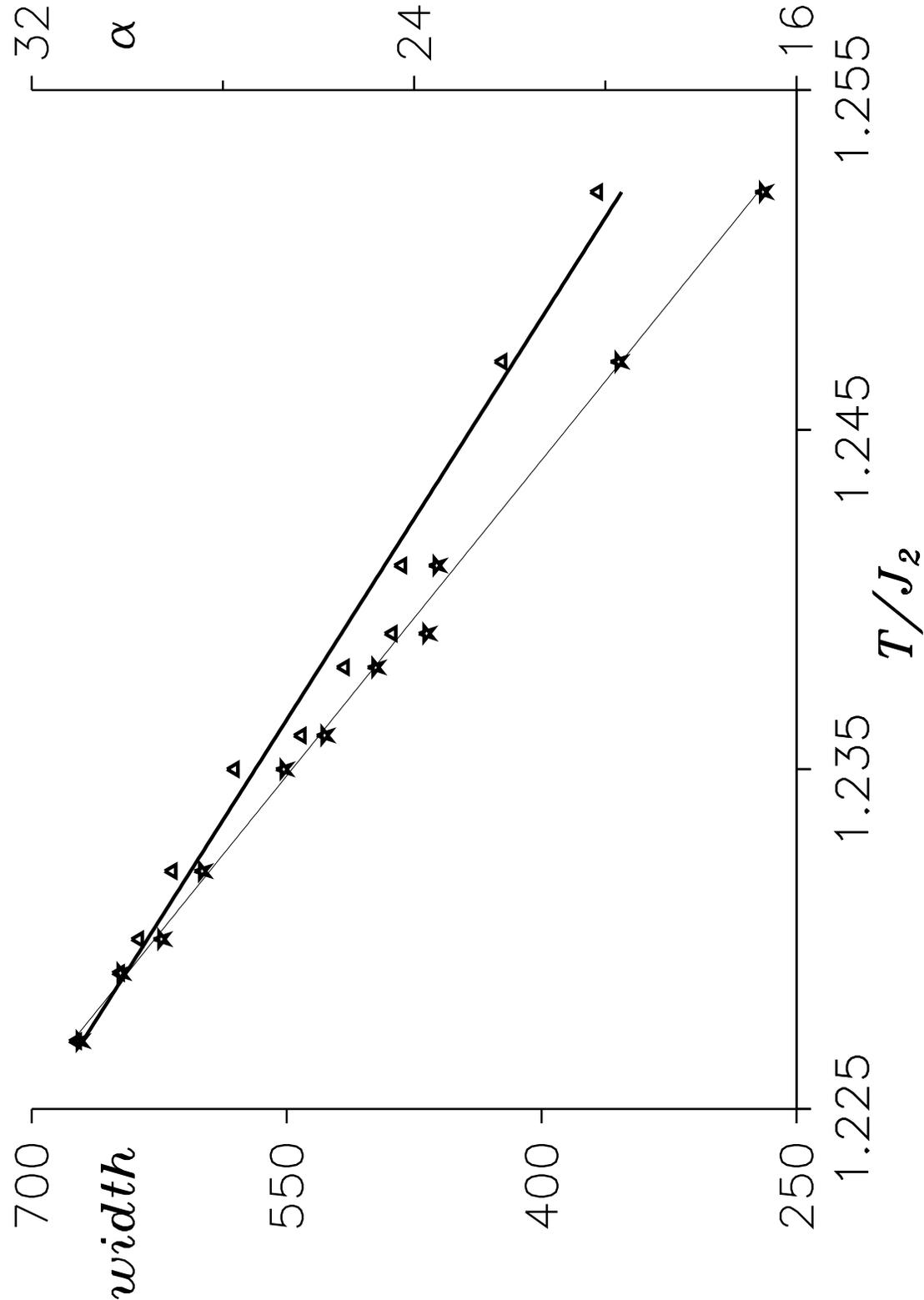
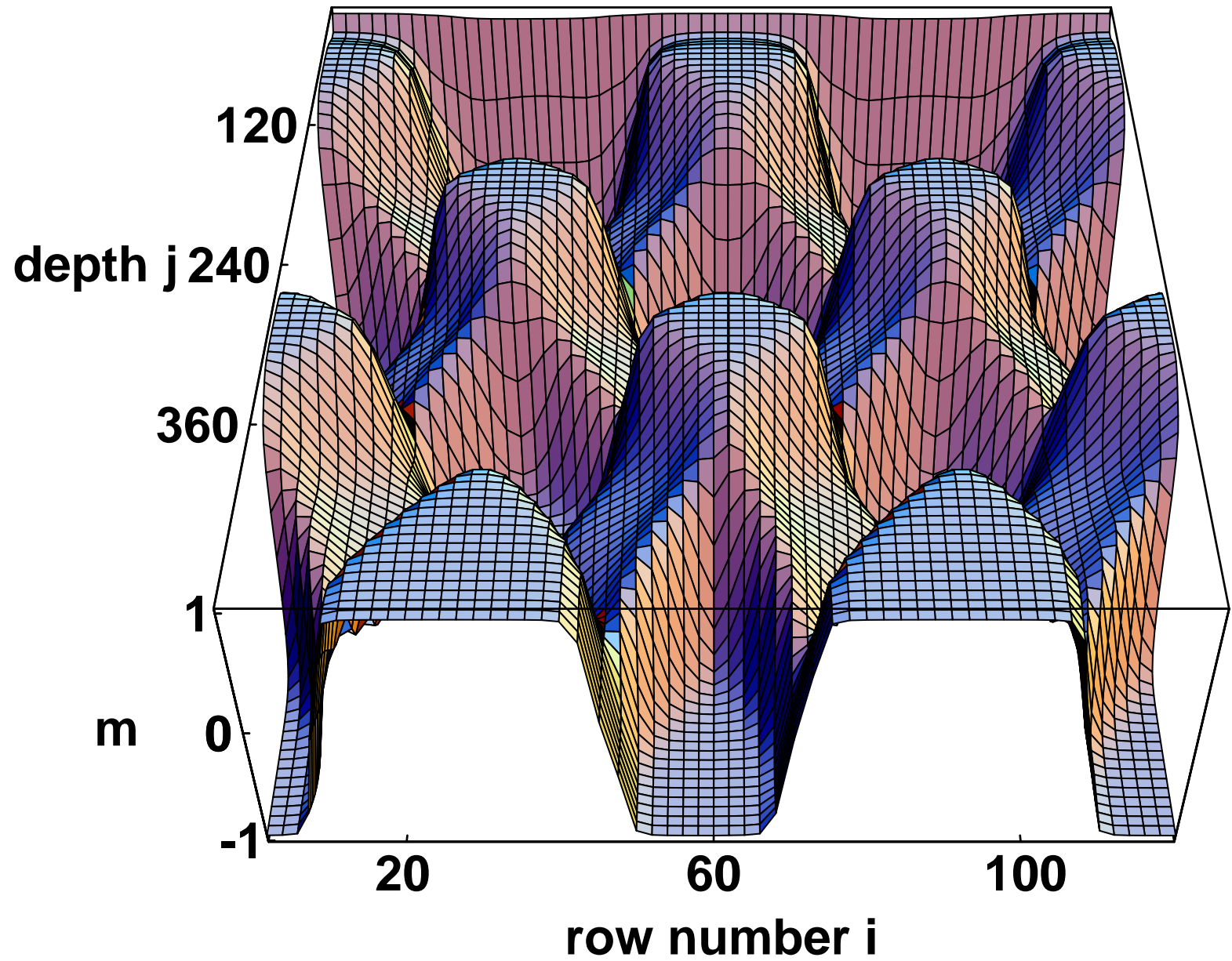
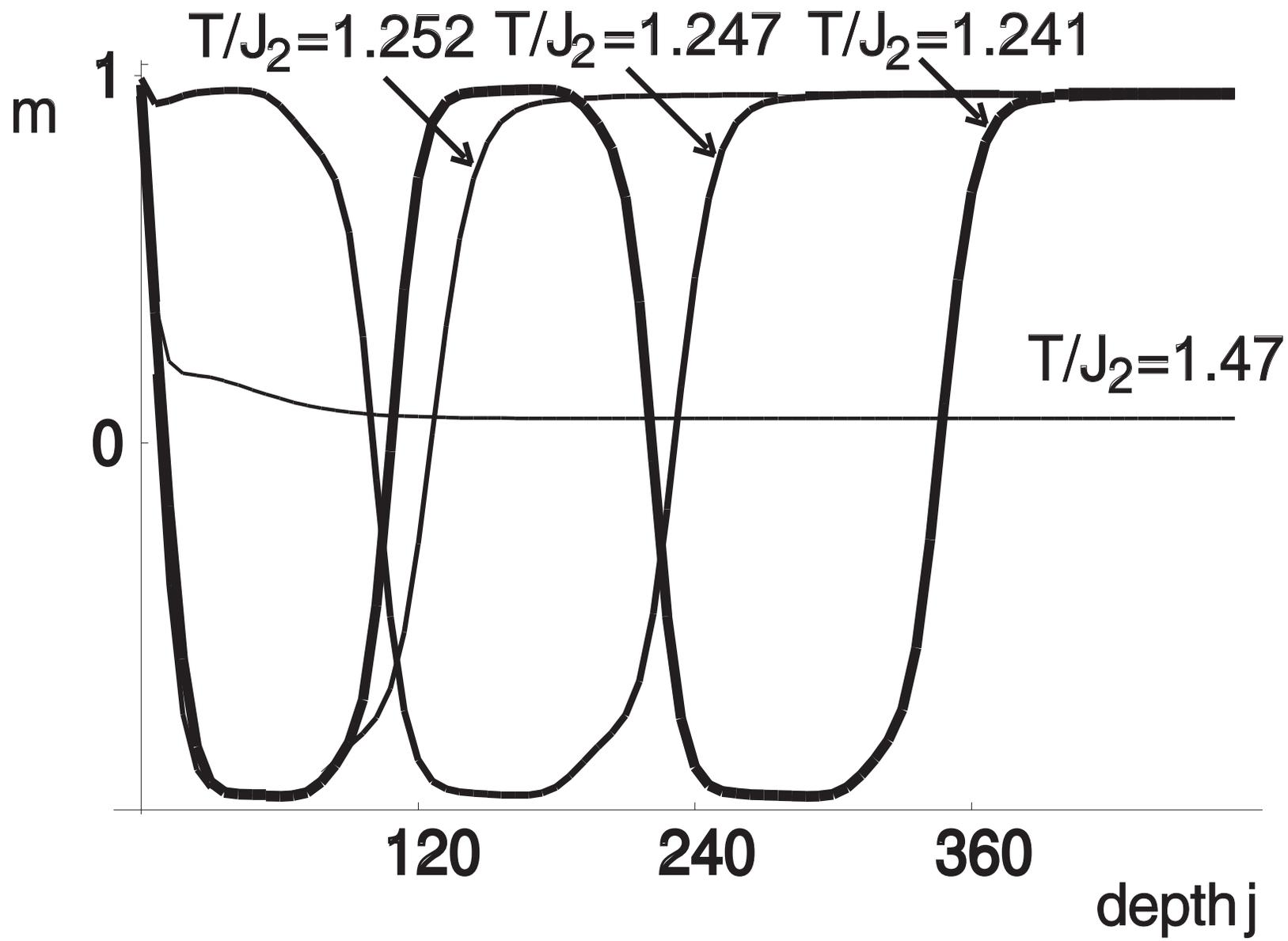
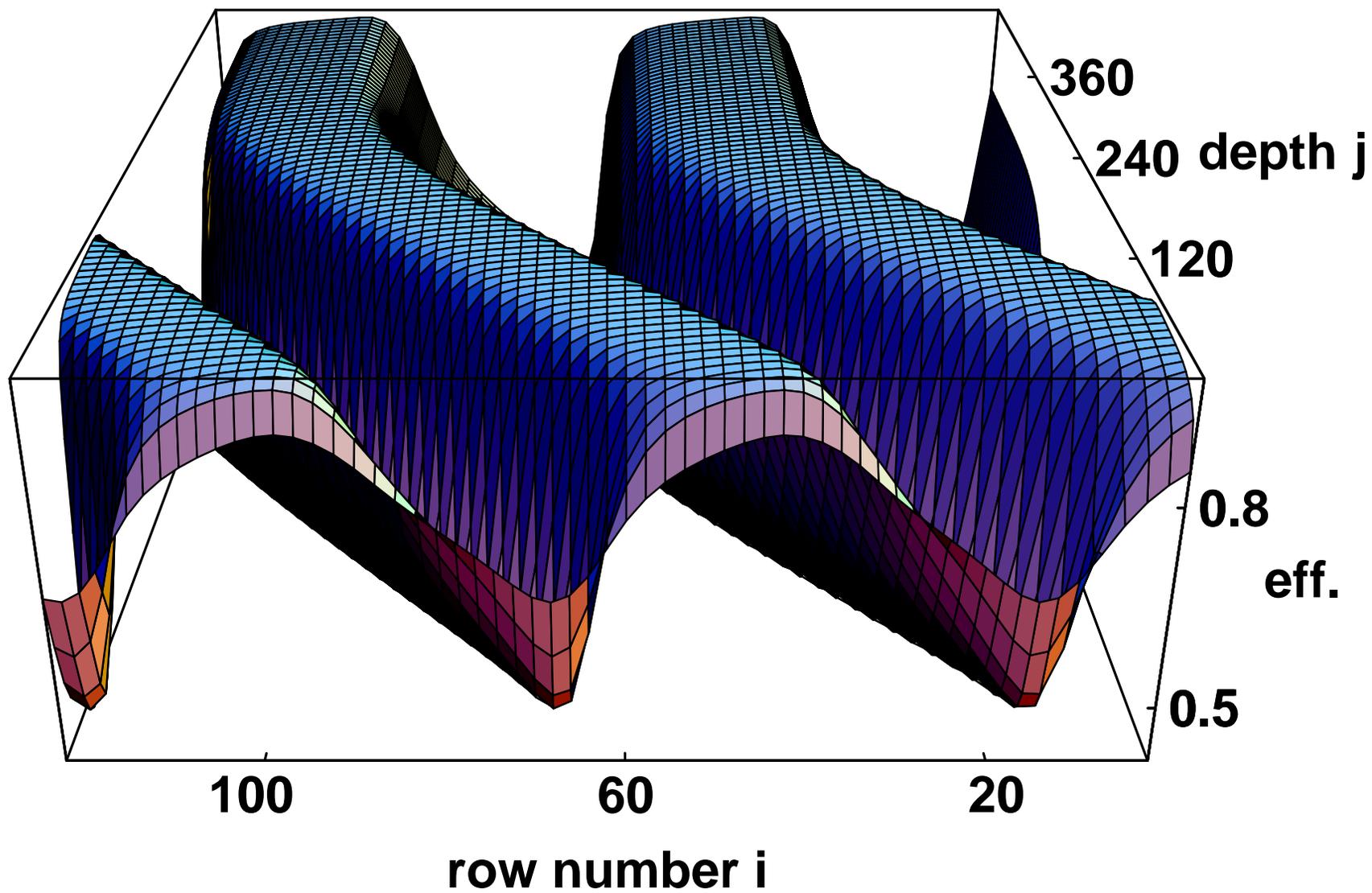
\bye